\documentclass{iaus}
\usepackage{graphicx}

\title[New class I methanol masers] 
{New class I methanol masers.}

\author[M.A.Voronkov \etal]   
{M. A. Voronkov$^{1,2}$, 
J. L. Caswell$^1$, S. P. Ellingsen$^3$, S. L. Breen$^1$, \\
T. R. Britton$^{4,1}$, J. A. Green$^{1}$, A. M. Sobolev$^{5}$ \and A. J. Walsh$^{6}$}

\affiliation{$^1$CSIRO Astronomy and Space Science, PO Box 76, Epping NSW 1710, Australia \\[\affilskip]
$^2$Astro Space Centre, Profsouznaya st. 84/32, 117997 Moscow, Russia \\[\affilskip]
$^3$School of Mathematics and Physics, University of Tasmania, GPO Box 252-37, Hobart, 
Tasmania 7000, Australia \\[\affilskip]
$^4$Macquarie University, Department of Physics and Engineering, NSW 2109, Australia \\[\affilskip]
$^5$Ural State University, Lenin ave. 51, 620083 Ekaterinburg, Russia \\[\affilskip]
$^6$Centre for Astronomy, School of Engineering and Physical Sciences, 
James Cook University, Townsville, QLD 4814, Australia}

\pubyear{2012}
\volume{287}  
\pagerange{119--126}
\jname{Cosmic Masers - from OH to H$_0$}
\editors{R.S. Booth, E.M.L. Humphreys \& W.H.T.Vlemmings, eds.}
\begin{document}

\maketitle

\begin{abstract}
We review properties of all known  collisionally pumped (class~I) methanol maser 
series based on observations with the  Australia Telescope Compact Array (ATCA)  and the 
Mopra radio telescope.  Masers at 36, 84, 44 and 95~GHz are most widespread, while 
9.9, 25, 23.4 and 104~GHz masers are much rarer, tracing the most energetic 
shocks. A survey of many  southern masers at 36 and 44~GHz suggests that these 
two transitions are highly complementary.
The 23.4 GHz maser is a new type of rare class~I methanol maser, detected only in 
two high-mass star-forming regions, G357.97-0.16 and G343.12-0.06, and
showing a behaviour similar to 9.9, 25 and 104~GHz masers.  
Interferometric positions suggest that shocks responsible for class~I masers could arise from
a range of phenomena, not merely an outflow scenario.
For example, some masers might be caused by interaction of an expanding H{\sc ii} region
with its surrounding molecular cloud. This has implications for evolutionary sequences 
incorporating class~I methanol masers if they appear more 
than once during the evolution of the star-forming region. We also make predictions
for candidate maser transitions  at the ALMA frequency range.
\end{abstract}

\firstsection 
\section{Introduction}

Methanol masers are associated with regions of active star formation,
with more than 20 different cm- and mm-wavelength
masing transitions discovered to date. The whole range of methanol
maser transitions does not share the same behaviour, loosely grouped in two classes. 
The division stems from early empirical distinctions
\cite[(e.g. Batrla et al. 1987)]{bat87}. Class~II  methanol 
masers (e.g. the most famous 6.7-GHz transition), along with OH and H$_{2}$O masers, occur
in the immediate environment of young stellar objects (YSOs) recognisable from their characteristic
infrared emission.  The class~II methanol masers are exclusive tracers of 
high-mass star-formation \cite[(e.g. Minier et al. 2003; Green et al. 2012)]{min03,gre12}. 
In contrast, the class~I masers (e.g. at 36 and 44 GHz), which are the subject of  
this paper, are usually found offset from the presumed origin of excitation
\cite[(e.g.  Kurtz et al. 2004; Voronkov et al. 2006)]{kur04,vor06}, and are 
found in regions of both high- and low-mass star formation 
\cite[(Kalenskii et al. 2010 and their paper in this volume)]{kal10}.

\begin{figure}[b]
\begin{minipage}{0.45\linewidth}
 \includegraphics[width=\linewidth]{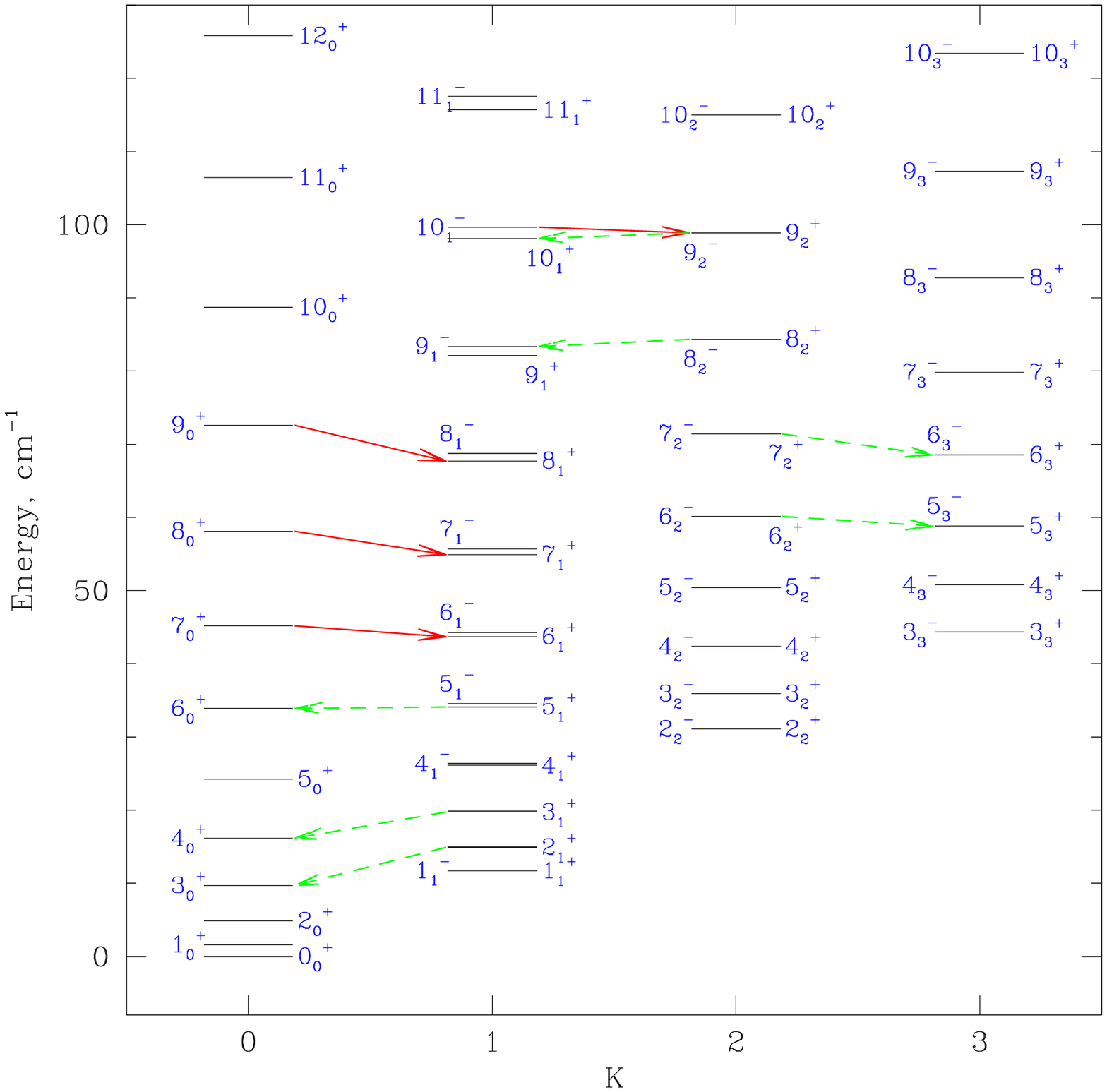} 
\end{minipage}\hfill
\begin{minipage}{0.45\linewidth}
 \includegraphics[width=1.0\linewidth]{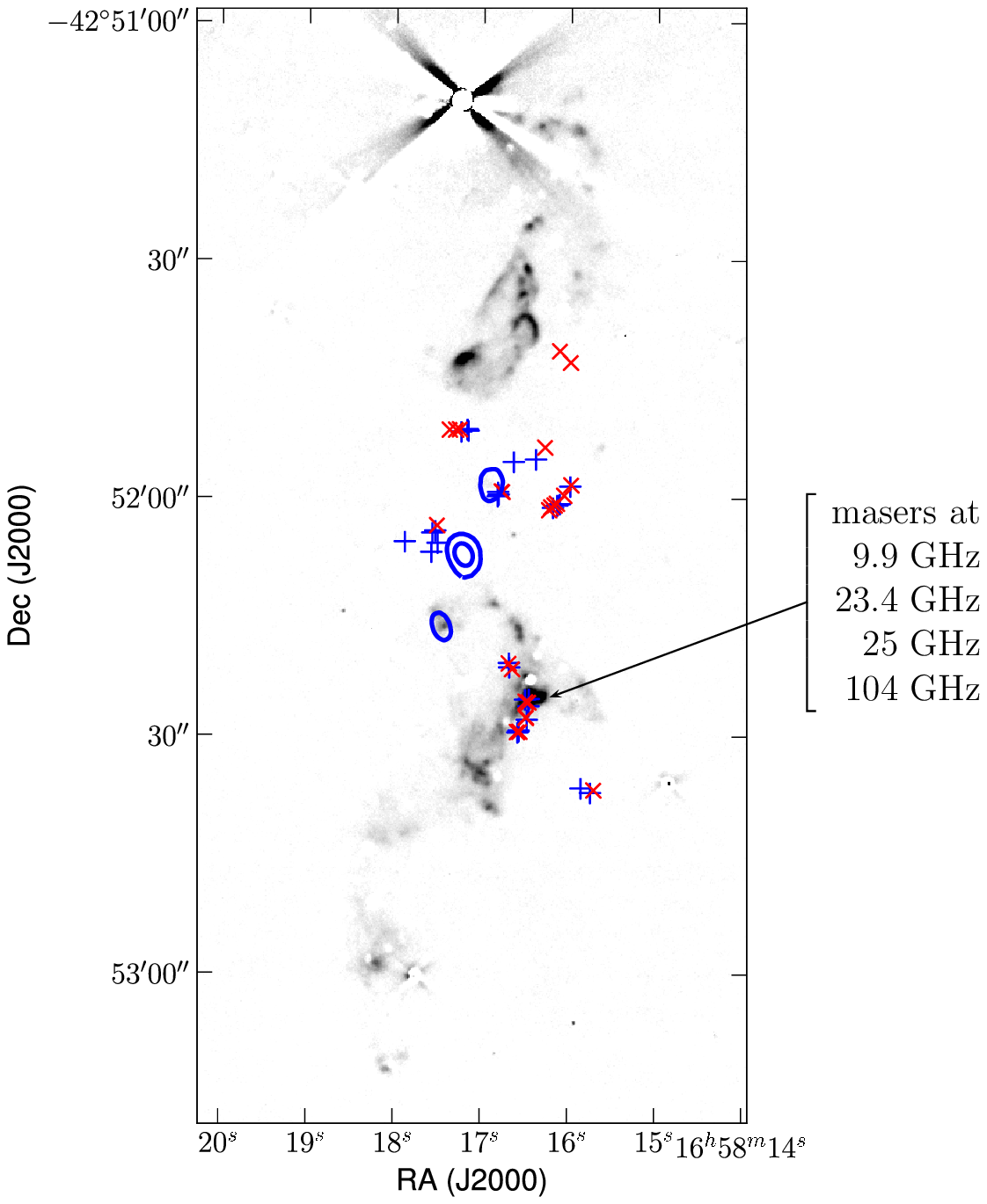} 
\end{minipage}
\begin{minipage}{0.45\linewidth}
 \caption{Energy level diagram for A-methanol (energies are given with respect to the lowest 
 level of A-methanol).  Solid (red) arrows represent known class~I 
 maser transitions, dashed (green) arrows show known class~II masers.}
 \label{meth_levels}
\end{minipage}\hfill
\begin{minipage}{0.45\linewidth}
 \caption{Distribution of the 36 (crosses) and 44 GHz (pluses) maser spots on top of the outflow image
 traced by 2.12$\mu$m  H$_2$ emission in G343.12-0.06, contours show the 12mm continuum 
 emission \protect\cite[(see also Voronkov et al. 2006)].}
 \label{g343.12}
\end{minipage}
\end{figure}

Theoretical calculations can explain 
this empirical classification, with the pumping process of 
class~I masers dominated by collisions (with molecular hydrogen), in contrast to 
class~II masers which are pumped by radiative excitation 
\cite[(e.g. Cragg et al. 1992; Voronkov 1999; Voronkov et al. 2005a)]{cra92,vor99,vor05a}.  
The two mechanisms are competitive \cite[(see Voronkov et al. 2005a for illustration)]{vor05a}: 
strong radiation from a nearby infrared source quenches class~I masers and 
strengthens  class~II masers.  The transitions of different 
classes occur in opposite directions between two given ladders of energy levels 
(Fig.~\ref{meth_levels}). The equilibrium breaks first between the ladders giving rise to either
class~I or class~II masers depending on whether the radiational or collisional excitation
dominate  \cite[(e.g. Voronkov 1999)]{vor99}. Therefore, bright masers of different classes 
residing in the same volume of gas are widely accepted as mutually exclusive (with potential
exceptions for weak masers).  However, on larger scales, they are often observed to 
coexist in the same star forming region within less than a parsec of each other (while
a few archetypal sources exist, displaying only one class of methanol maser).

In addition to gross classification, there are finer distinctions within the same class of methanol
maser transitions. At sensitivity levels typically attained in surveys, the range of transitions
can be further categorised into widespread masers (e.g. at 44~GHz) and rare or weak 
masers (e.g. at 9.9~GHz). Models seem to suggest that
rare masers require higher temperatures and densities to form (Sobolev et al. 2005).
The maser transitions of methanol tend to form series (individual
transitions have different quantum numbers J as evident from Fig.~\ref{meth_levels}). Observational properties such as whether the individual transitions give rare or widespread masers
are qualitatively similar within the same series. Superposed are trends with J caused by the 
changes of excitation energy and the efficiency of the sink process due to a different number
of energy levels below. Interestingly, all class~II methanol maser 
series  (with the exception of J$_2$-(J-1)$_3$~A$^{\pm}$ series based on the 38-GHz maser) are 
going downwards (with J decreasing while frequency increases) and eventually terminate. 
In contrast, all class~I maser series extend upwards (see Fig.~\ref{meth_levels}). Therefore,
the majority of candidate maser transitions searchable with the Atacama Large Millimetre Array
(ALMA) in the millimetre and sub-millimetre bands belong to class~I. In the following sections
we review observational properties of all known class~I methanol maser series before summarising
predictions for ALMA in bands 6 and 7. For simplicity, we refer to the maser series by the
lowest frequency transition.

\section{Widespread class~I masers (series based on 36 and 44 GHz masers)}

The J$_0$-(J-1)$_1$~A$^+$ methanol series includes the most studied 44 and 95-GHz 
class~I methanol masers. A few hundred such masers are currently known, but the majority 
have only single dish data \cite[(e.g. Haschick et al. 1990; Slysh et al. 1994; 
Val'tts et al. 2000; Ellingsen 2005; Fontani et al. 2010; Chen et al. 2011; 
unpublished Mopra data from our group)]{has90,sly94,val00,ell05,fon10,che11}. The major
published interferometric surveys are those of \cite[Kurtz et al. (2004)]{kur04} and 
\cite[Cyganowski et al. (2009)]{cyg09}. The second class~I maser series in the widespread 
category is  J$_{-1}$-(J-1)$_0$~E which is renowned for masers at 36 and 84~GHz. These two
maser transitions are considerably less studied than the 44 and 95-GHz pair. As before,
most observational data are obtained with single dish facilities \cite[(e.g. Haschick \& Baan 1989;
Kalenskii et al. 2001)]{has89, kal01}. The reported interferometric observations are scarce
and confined to single source papers only \cite[(e.g. Voronkov et al. 2006; 
Voronkov et al. 2010; Sjouwerman et al. 2010; Fish et al. 2011)]{vor06,vor10a,sjo10,fis11}.
The typical spread of maser spots is comparable to or exceeds the beam size of a 20-m 
class single dish at the frequencies of these transitions \cite[(Kurtz et al. 2004; Voronkov et al. 2006)]{kur04,vor06}. Therefore, interferometric observations, which allow us to measure positions
of each maser sport accurately, are crucial even to get meaningful detection statistics.

To increase the number of class~I masers studied at high 
angular resolution and to compare the morphologies observed in different maser transitions 
we carried out in 2007 a quasi-simultaneous interferometric survey at 36 and 44~GHz of 
all class~I masers reported in the literature at the time of the observations and located south of declination $-$35$^o$ \cite[(a single source from the project was presented in Voronkov et al. 2010a)]{vor10a}.
Fig.~\ref{g343.12} shows the results of this survey for G343.12-0.06, which has been studied in 
detail in other transitions by
\cite[Voronkov et al. (2006)]{vor06}. 
The distribution of 36 and 44~GHz maser spots resembles that of 84- and 95-GHz maser spots
from \cite[Voronkov et al. (2006)]{vor06}, which is a good example of outflow association, but also 
has a few new spots found due to the larger primary beam and higher signal-to-noise ratio of the 
new observations. Note, that all rare 
masers in this source are located at the same position near the brightest 
knot of the 2.12~$\mu$m molecular hydrogen emission, which is a well known shock tracer
\cite[(Voronkov et al. 2006)]{vor06}. With the caveat about extinction variations, this supports the 
idea that rare class~I masers require higher temperatures and densities to form than the 
widespread masers \cite[(Sobolev et al. 2005)]{sob05}. 

\begin{figure}[b]
\begin{minipage}{0.45\linewidth}
 \includegraphics[width=1.0\linewidth]{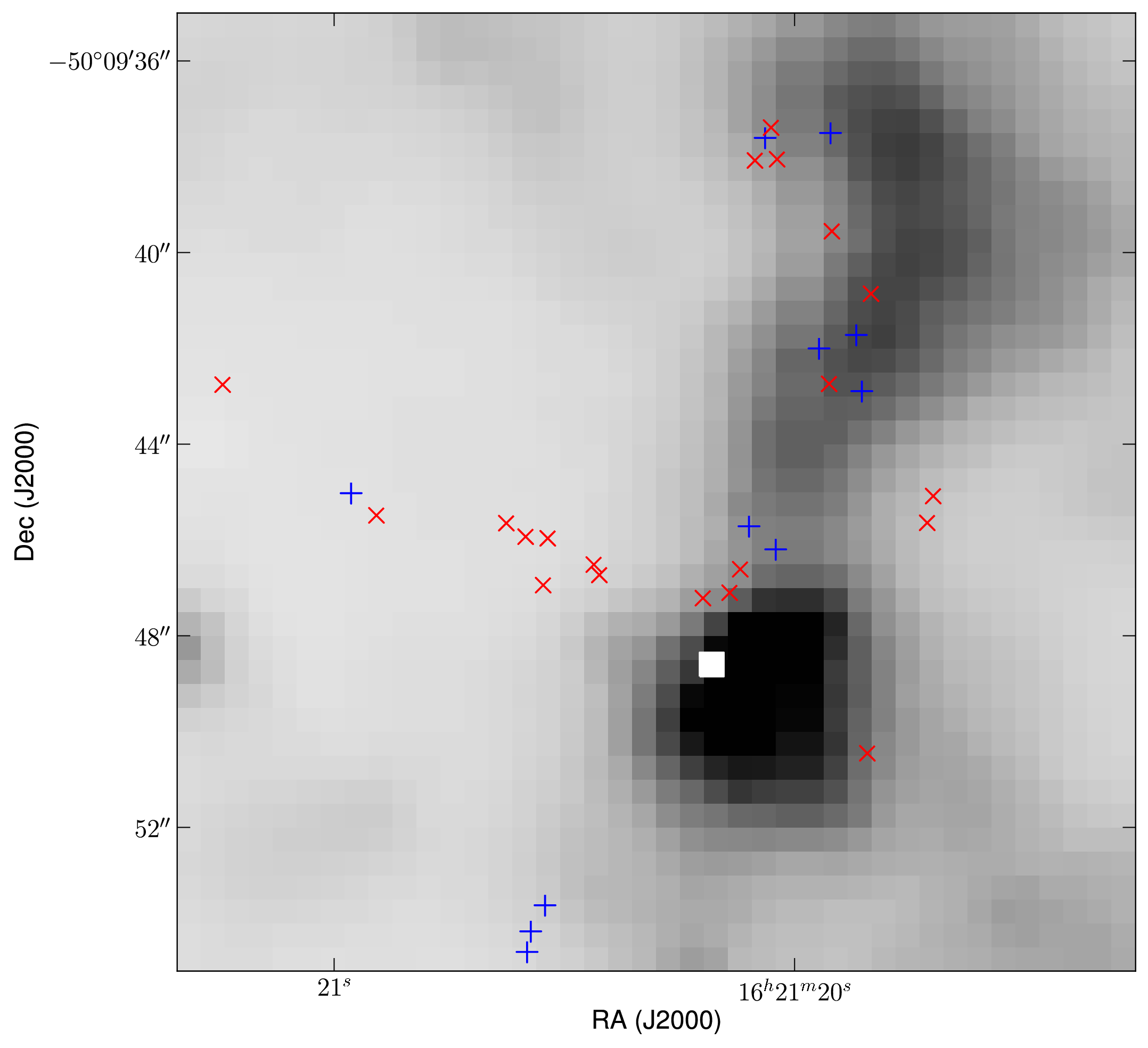} 
\end{minipage}\hfill
\begin{minipage}{0.45\linewidth}
\hskip -3mm
 \includegraphics[width=1.05\linewidth]{mvoronkov_fig4.eps} 
\end{minipage}
\begin{minipage}{0.45\linewidth}
 \caption{Distribution of the 36 (crosses) and 44 GHz (pluses) maser spots in G333.466-0.164. 
 The position of the 6.7-GHz maser is shown by filled square. The background is GLIMPSE
 4.5-$\mu$m image.}
 \label{g333.466}
\end{minipage}\hfill
\begin{minipage}{0.45\linewidth}
\caption{The position of the 9.9-GHz methanol maser in W33-Met. The contours represent
8.4-GHz continuum image, grayscale shows the distribution of the thermal NH$_3$ emission 
\protect\cite[(for details see Voronkov et al. 2010b and references therein)]{vor10b}.}
\label{w33met}
\end{minipage}
\end{figure}

In G343.12-0.06, the majority of 44-GHz maser spots have some 36-GHz emission and vice 
versa (Fig.~\ref{g343.12}). However, in many cases these two transitions were found to be
highly complementary. Fig.~\ref{g333.466} shows the maser spot distribution in G333.466-0.164,
the best example of such a scenario that we currently have. The 44-GHz maser spots are distributed 
roughly along the line traced by the source of extended infrared emission with 4.5-$\mu$m excess
(often referred to as an Extended Green Object or EGO). Without the 36-GHz data, this
EGO would most likely be interpreted as tracing an outflow emanating from the location of the 
YSO marked by the 6.7-GHz maser (shown by square in Fig.~\ref{g333.466}). The chain 
of 36-GHz maser spots completes the second half of a bow-shock structure suggesting a 
different direction of the outflow. Another good example is the high-velocity feature blue-shifted by 
about  30~km~s$^{-1}$ from the systemic velocity which was found in G309.38$-$0.13 at 36-GHz only 
\cite[(Voronkov et al. 2010a)]{vor10a}.  It is worth noting, that \cite[Sobolev et al. (2005)]{sob05}
suggested that the 36 to 44-GHz flux density ratio is very sensitive to the orientation of the maser 
region.

\section{Rare 9.9 and 104 GHz masers}
These masers belong to the J$_{-1}$-(J-1)$_{-2}$~E methanol series, J=9 and 11, respectively. 
The first search for 9.9-GHz masers was carried out by \cite[Slysh et al. (1993)]{sly93} who reported a 
single maser detection towards W33-Met (G12.80$-$0.19). Recently, \cite[Voronkov et al. (2010b)]{vor10b} 
carried out  a sensitive (1$\sigma$ limits as low as 100~mJy) survey at 9.9-GHz with the ATCA and
found 2 new detections out of 46 targets observed. Two additional 9.9-GHz masers in G343.12$-$0.06 
and G357.97$-$0.16 were found serendipitously \cite[(Voronkov et al. 2006, 2011)]{vor06,vor11}.
The latter maser is the strongest, with peak flux density around 70~Jy and the only one
for which the absolute position has not been measured (although the position is expected to be 
the same as for the 23.4-GHz maser found in this source). 

With the exception of the 104-GHz maser in G343.12$-$0.06 which had ATCA observations \cite[(Voronkov et al. 2006)]{vor06}, all other currently known 104-GHz masers were found using 
single dish facilities \cite[(Voronkov et al. 2005b, 2007)]{vor05b,vor07}. 
In addition to the sources of 9.9-GHz maser emission described above, these observations
 brought only one new maser in G305.21$+$0.21. It is worth noting that a weak 
maser at 9.9~GHz was seen towards this source
during test ATCA observations, but happened to be below the detection threshold of the regular 
survey \cite[(Voronkov et al. 2010b)]{vor10b}. This brings the total number of known masers
in this series to 6, in contrast to more than 200 hundred known widespread masers.

Detailed investigations of these masers suggests that some class~I masers (in all
class of transitions, not just 9.9-GHz) may be caused by expanding HII regions 
\cite[(see e.g. Fig.~\protect\ref{w33met} and Voronkov et al. 2010b)]{vor10b}. This is an
additional scenario to the commonly accepted mechanism for the formation of class~I 
masers in the outflow shocks. 

\section{Evolutionary stage of star-formation with class~I masers}
The question whether different 
masers trace distinct evolutionary stages of high-mass star formation has recently become 
a hot topic \cite[(see e.g. Breen et al. 2010 and references therein)]{bre10}, although the place
of class~I masers in this picture is still poorly understood. \cite[Ellingsen (2006)]{ell06} 
investigated the infrared colours of GLIMPSE (Galactic Legacy Infrared Mid-Plane Survey 
Extraordinaire) catalogue  point sources associated with methanol masers and suggested
that class~I methanol masers may signpost an earlier stage of high-mass star formation than
the class~II masers. These and other considerations laid the foundation of a
qualitative evolutionary scheme for different maser species proposed by  
\cite[Ellingsen et al. (2007)]{ell07}. The scheme was further refined by 
\cite[Breen et al. (2010)]{bre10} in their Figure 6, but with no new survey data available on 
class~I masers  the conclusion about evolutionary stage when these masers are present essentially
remained the same. \cite[Voronkov et al. (2006, 2010b)]{vor06,vor10b} pointed out that this
statement is inconsistent with detailed studies of class~I maser sources 
which do overlap with OH masers. Moreover, we carried out a search for 
the 44-GHz class~I methanol masers towards known OH masers which were not detected
at 6.7-GHz (class~II) in the unbiased Methanol Multi-Beam (MMB) survey \cite[(Green et al. 2012 and references therein)]{gre12}. Despite an inadequate spectral resolution (about 7~km~s$^{-1}$) 
achieved in these test observations, which makes the survey insensitive to weak ($<$15~Jy) masers,
the detection rate exceeded 50\%. Therefore, it seems more appropriate to place the class~I 
methanol masers as partly overlapping, but largely post-dating the evolutionary phase associated 
with class~II methanol masers. The suggested association with  expanding HII regions 
also implies that some of the sources are quite evolved. Whether there is a population of
class~I masers pre-dating the phase with the class~II masers is unclear at present 
\cite[(see Voronkov et al. 2010b and Chen et al. 2011 for detailed discussion of these issues)]{vor10b,che11}.

It is also important to keep in mind the major assumptions which underlie the 
evolutionary timeline suggested by \cite[Ellingsen et al. (2007) and Breen et al. (2010)]{ell07,bre10}.
In particular, that each major maser species arises only once during the evolution of a 
particular star formation region, and that all the maser species are associated with a 
single astrophysical object. The large spatial (and angular) spread of class~I maser spots
makes such identifications much less clear given a generally crowded environment of high-mass star 
formation where sequential or triggered star formation may take place. It is possible that
one YSO in the cluster may have an associated 6.7-GHz maser, while another
could be associated with OH masers and a well developed HII region.  At this
stage the number of sufficiently simple sources studied in detail is quite limited 
hindering the statistical analysis and even in relatively simple sources some controversy may
exist (see e.g. the case of G357.97$-$0.16 discussed by Britton et al., this volume).
It is worth noting that the survey of \cite[Chen et al. (2011)]{che11} supports the hypothesis that 
class~I masers may arise at more than one evolutionary phase.

\section{Weak 25-GHz series}
Historically, the first methanol masers found in space were from the J$_2$-J$_1$~E class~I methanol maser series near 25-GHz towards Orion \cite[(Barrett et al. 1971)]{bar71}. Very few
additional sources were found in the following 3 decades, so these masers were widely believed
to be rare. \cite[Voronkov et al. (2007)]{vor07} searched for J=5 transition of this maser series
towards 102 targets with ATCA (although no absolute position was measured during this experiment).
This search yielded 66 detections, but mostly weaker than 1~Jy. Other recent studies
\cite[(e.g. Brogan et al. 2011; paper in this volume)]{bro11} also suggest that these masers are
common, but typically weak. The new ATCA backend enabled simultaneous observations
of up to 9 (limited by  the receiver frequency range) transitions of this series \cite[(e.g. Wilson et al. (2011)]{wil11}. Britton \& Voronkov (paper in this volume) have recently followed up 
the majority of known southern 25-GHz masers in the J=2 to J=9 transitions.

\section{New rare class~I methanol maser at 23.4-GHz}
The H$_2$O southern Galactic Plane Survey \cite[(HOPS; Walsh et al., 2011 and their paper in this volume)]{wal11} brought
an unexpected discovery of a new class~I methanol maser at 23.4-GHz in G357.97$-$0.16 
\cite[(see Voronkov et al. 2011 for details)]{vor11}. This is the first transition in the 
J$_1$-(J-1)$_2$~A$^-$ class~I maser series (corresponding to J=10). A weaker
23.4-GHz maser was found in G343.12$-$0.06. Observational properties of this new maser are 
similar to other rare masers like that at 9.9-GHz.
The source of strongest 23.4 and 9.9-GHz maser emission, G357.97$-$0.16, is discussed
in detail by Britton et al. (this volume).

\section{Class~I methanol masers at high spatial resolution}
In contrast to  class~II methanol masers, Very Long Baseline Interferometry (VLBI) observations 
of class~I masers were effectively abandoned  following a few unsuccessful attempts in 
early 1990s (which remained unpublished).  The understanding of class~I masers has certainly 
progressed since then and observations with connected element interferometers have become 
available  \cite[(e.g.  Kurtz et al. 2004;  Cyganowski et al. 2009; see also earlier 
sections)]{kur04,cyg09}. In hindsight, the reasons for failure to detect
class~I masers in early VLBI experiments were not limited to larger intrinsic sizes of maser
spots and
 a general difficulty doing high-frequency VLBI. First, only single dish positions (accurate up to arcmin) 
were available for most if not all of the targets. It is shown in the previous sections that
images of class~I masers typically contain several spots distributed over a large area often 
covering the whole primary beam of the telescope 
\cite[(see also Voronkov et al. 2006; Kurtz et al. 2004)]{vor06,kur04}. Therefore, 
there are high chances to miss a spot of emission given a typical narrow field of view achieved in
a VLBI experiment. In addition, the lack of images at arcsecond resolution made the selection of 
best most compact targets difficult. Recent VLBI 
observations (Tarchi, priv. comm.) at 44-GHz revealed fringes at the shortest baselines (similar observations were also carried out by Kim et al. (priv. comm.) with the Korean VLBI Network). 
The second issue is the
selection of maser transition. The early VLBI attempts targeted the strongest and most
widespread maser transitions at 44 and 95-GHz transitions. However, these are likely to 
have larger intrinsic source sizes being easier to excite 
than, for example, masers at 9.9 and 23.4~GHz along with the masers in the series 
near 25~GHz. It is worth noting that VLBI observations of 10 strongest 25-GHz masers have 
recently been carried out with the Long Baseline Array (LBA).

\section{Millimetre and sub-millimetre masers with ALMA}
 One can extend the maser series reviewed in the previous sections to higher frequencies.
 ALMA bands 6 and 7 (already implemented) encompass the following candidate maser transitions. 
The series based on the 36-GHz maser giving 8$_{-1}$-7$_0$~E at 229~GHz 
\cite[(a known maser: Slysh et al. 2002; Fish et al. 2011)]{sly02,fis11},  
9$_{-1}$-8$_0$~E at 278~GHz recently found in S255N with the 
SMA (Salii, Sobolev, Zinchenko, Liu \& Su, priv. comm) and 
10$_{-1}$-9$_0$~E at 327~GHz, although the
system performance is poor for the latter transition. The series based on the 44-GHz maser giving
11$_0$-10$_1$~A$^+$ at 250~GHz, 12$_0$-11$_1$~A$^+$ at 303~GHz and 13$_0$-12$_1$~A$^+$ at 356~GHz. The series based on the 9.9-GHz maser giving 14$_{-1}$-13$_{-2}$~E at 242~GHz,
15$_{-1}$-14$_{-2}$~E at 287~GHz, 16$_{-1}$-15$_{-2}$~E at 331~GHz. The new series
based on the 23.4-GHz maser giving 14$_1$-13$_2$~A$^-$ at 237~GHz, 15$_1$-14$_2$~A$^-$ at 291~GHz and 16$_1$-15$_2$~A$^-$ at 346~GHz. 

The field of high-frequency class~I methanol masers is essentially uncharted territory. For possible
masers at even higher frequencies (e.g. ALMA band 9), it is hard to make sensible
predictions. The transitions listed above correspond to excitation
energies of about 300-500~K. The 25-GHz transitions are a cm-wavelength series but, between the same two ladders of levels, is the  J$_2$-(J-1)$_1$~E series which gives mm-wavelength transitions
including a 218-GHz maser (J=4) detected in the SMA observations mentioned above. It is worth noting that maser models predict population inversion for all these transitions (see also Sobolev
et al., this volume).
 
\section{Conclusions}
\begin{enumerate}
\item Studies of different maser transitions are very complementary (filling the dots in
morphology, high-velocity features, modelling). 
\item Rare/weak class~I methanol masers (9.9, 23.4, 25 and 104~GHz) trace stronger
shocks and higher temperatures and densities.
\item Some class~I masers may be caused by expanding HII regions, a scenario additional
to their formation in outflows. An implication for a maser-based evolutionary sequence is that
class~I methanol masers may appear more than once during YSO evolution,  and thus some 
regions with class~I masers are probably quite evolved.
\item The evolutionary stage with class~I masers probably outlasts the stage
when the 6.7-GHz methanol masers (class~II) are present,  overlapping with OH 
maser  activity.
\item Promising ALMA maser targets are G343.12-0.06 and G357.97-0.16. 
\end{enumerate}

\section*{Acknowledgments}
The Australia Telescope Compact Array and the Mopra telescope are
parts of the Australia Telescope National Facility which is funded by the
Commonwealth of Australia for operation as a National Facility managed
by CSIRO.  The University of New South Wales Digital Filter Bank used for the 
observations with the Mopra Telescope was provided with support from the 
Australian Research Council. The research has made use of the NASA/IPAC Infrared
Science Archive, which is operated by the Jet Propulsion Laboratory,
California Institute of Technology, under contract with the National
Aeronautics and Space Administration. AMS would like to thank the Russian 
Foundation for Basic Research (grant 10-02-00589-a) and the Russian federal task 
program ``Research and operations on priority directions of development of the 
science and technology complex of Russia for 2007-2012'' (state contract 16.518.11.7074).

\end{document}